\documentclass[]{spie}  %>>> use for US letter paper
\usepackage{amsmath,amsfonts,amssymb}
\usepackage{graphicx}
\usepackage{subfigure}
\usepackage[colorlinks=true, allcolors=blue]{hyperref}
\usepackage{svg}
\usepackage{gensymb}
\usepackage{amssymb}
\usepackage{wasysym}
\usepackage{textcomp}
\usepackage{tcolorbox}
\usepackage{svg}

\title{A Multi-Functional Fiber Positioning System for Extremely Large Telescopes}

\author[a,b]{Manjunath Bestha}
\author[a]{T. Sivarani}
\author[a]{Arun Surya}
\author[a]{Sudharsan Yadav}
\author[c]{Athira Unni}
\author[a,b]{Parvathy M}
\author[a]{Devika Divakar}
\author[a]{S. Sriram}
\author[a]{Ajin Prakash}
\author[a]{Amirul Hasan}
\affil[a]{Indian Institute of Astrophysics, India}
\affil[b]{University of Calcutta, India}
\affil[c]{University of California, Irvine, USA}

\authorinfo{Further author information(Send correspondence to Manjunath): \\ Manjunath B: E-mail: bestha95@gmail.com\\  Sivarani T: E-mail: sivarani@gmail.com}

\begin{document} 

\maketitle

\begin{abstract}

We present a conceptual design for a fiber positioning system for multi-object high-resolution spectroscopy, designed to be compatible with the upcoming large telescopes with a wide field of view. The design incorporates multiple Atmospheric Dispersion Correctors (ADCs) and tip-tilt mirrors that receive non-telecentric input from individual targets and direct it to the ADCs. Here, we introduce a mechanical design for the fiber positioner that accommodates the optics and operates in a curved focal plane with a Radius of Curvature (R) of 3m. This mechanical design provides four degrees of freedom to access the focal volume, enhancing targeting efficiency. The proposed design and an efficient target allocation algorithm ensure a targeting efficiency of approximately 80-100\% for a primary observation session. We also present a methodology for target assignment, positioning, and quantification based on sequential and Monte Carlo (MC) algorithms. This method has been tested on realistic fields with varying target densities to validate its performance.

\end{abstract}


\section{Introduction}

Fiber-fed multi-object high-resolution spectroscopy is quite limited among large telescopes. This limitation arises because high-resolution spectroscopy is photon-starved and requires a wide field to choose suitable bright targets. However, wide-field spectroscopy becomes more challenging for larger telescopes due to a curved focal plane, non-telecentricity, and their large plate scale, which then requires a large Atmospheric Dispersion Corrector (ADC). These challenges can be addressed; for instance, a small atmospheric dispersion corrector can be placed in a fiber positioner arm \cite{zebri_2014}, and non-telecentricity can be corrected with a combination of a pick-off mirror and collimator.

Therefore, multi-functional fiber positioners are needed to accomplish these tasks. One such positioner is the K-band Multi-object Spectrograph (KMOS), which, although not fiber-fed, operates with two degrees of freedom (DOF) by selecting narrow fields using robotic pick-off arms. These arms pivot around the periphery of the focal plane of the Very Large Telescope (VLT), can rotate about its axis ($\phi$), and radially move in and out (\textit{r}) as shown in Figure \ref{Dof}. Each KMOS positioner hosts the Integral Field Unit (IFU) and feeds light to the spectrograph. Considering KMOS's design and functioning as a multi-functional fiber positioning system \cite{sharples_2013}, we propose a new concept of Fiber Positioning System (FPS) with four DOFs by adding two DOFs to the KMOS design (see Section \ref{section1}). The advantages of these additional DOFs are their potential use for surveys and observing specific targets for specific scientific objectives. For instance, in multi-object transmission spectroscopy, where both the program star and the reference star are observed simultaneously, which currently relies on slit-based low-resolution spectroscopy, can be extended to a fiber-fed system.

In this paper, we present the conceptual design and specific methodology for target assignment and placement of positioner's ferrule on targets in a Multi-object Spectroscopic Mode (MOS) of the High-Resolution Optical Spectrograph (HROS) of the Thirty Meter Telescope (TMT) \cite{Sivarani_2022}. This concept is beneficial for seeing limited spectrographs on telescopes with larger plate scales and having a large image size on the focal plane, requiring less precision in positioning. For instance, TMT has a plate scale of 2.18 mm per arcsecond \cite{Sivarani_2022}, and due to atmospheric dispersion, the image size can increase to about 5" ($\approx$ 10mm) at a zenith angle of 60\degree on TMT's focal plane\cite{Manju_2023}. Given that the precision of current actuators can reach the micrometer level, the accuracy of positioning the ferrule on a 10mm image may not pose a significant challenge for this concept.
%It varying from 2mm at zenith 0$\degree$ to 10mm at zenith 60$\degree$.

\label{intro}  % \label{} allows reference to this section

\section{The Mechanical Concept}
\label{section1}

\begin{figure}[h!]
    \centering
    \subfigure{\includegraphics[width=1\textwidth]{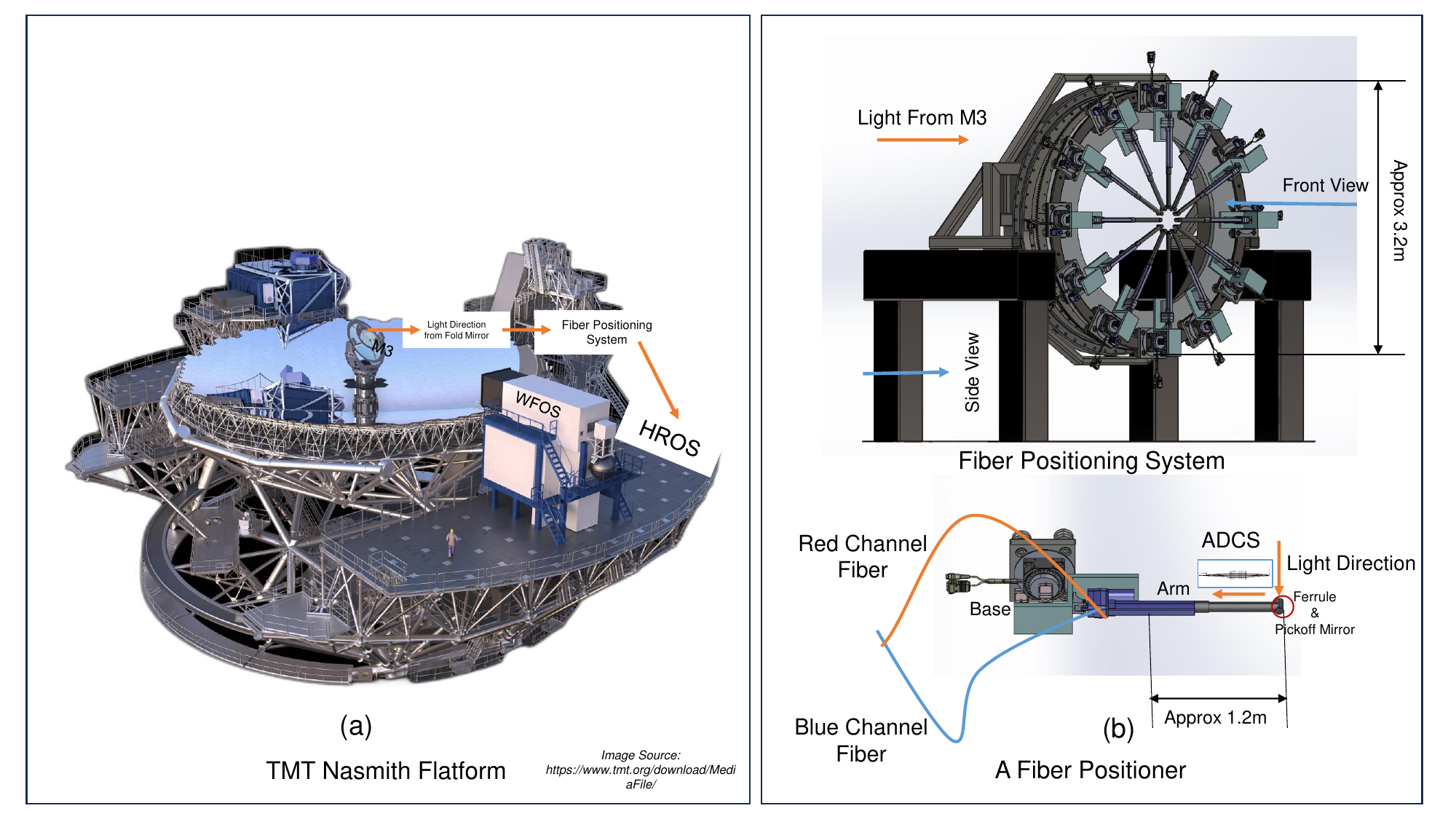}}
    \caption{The mechanical schematic showing the location of the positioning system on the telescope's Nasmyth platform and the concept design of the positioning system and positioner.}
    \label{Mech_Design}
    
    \subfigure{\includegraphics[width=1\textwidth, trim=0pt 175pt 0pt 140pt, clip]{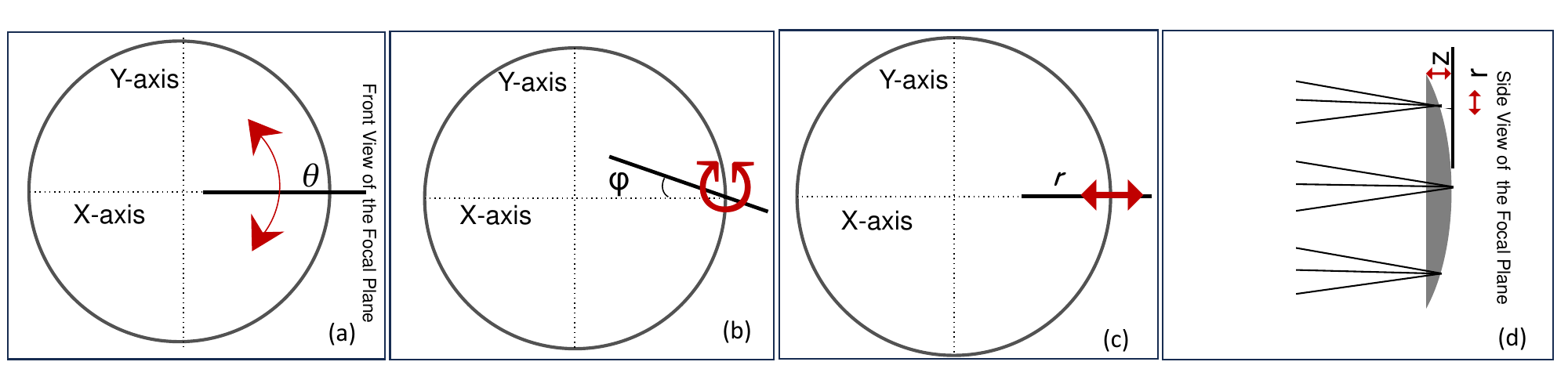}}
    \caption{The positioner's degrees of freedom: the circle represents the focal plane's front view, the black line segment is the positioner arm, and the arrows indicate the direction of movement. (a) \(\theta\) rotational motion around the focal plane, (b) \(\phi\) rotational motion around the positioner pivot, (c) \(r\) linear motion in the radial direction, and (d) \(z\) linear motion in the axial direction. The three sets of rays represent field objects at the focal plane.}
    \label{Dof}
\end{figure}

The concept of a positioning system involves a collection of fiber positioners, each consisting of three main parts: the base, arm, and ferrule, as shown in Figure \ref{Mech_Design}b. These positioners are arranged around the focal plane on a field rotator that rotates the entire system to correct for field rotation. This system will be located on the Nasmyth platform of the telescope, in front of the spectrograph, as depicted in Figure \ref{Mech_Design}a. Light from the telescope's fold mirror (M3) will be diverted to the positioning system, where individual fiber positioners collect light from specific objects (i.e., the field corresponding to the diameter of the pick-off mirror) on the focal plane.

\begin{figure}[h!]
    \centering
    \subfigure{\includegraphics[width=1.02\textwidth]{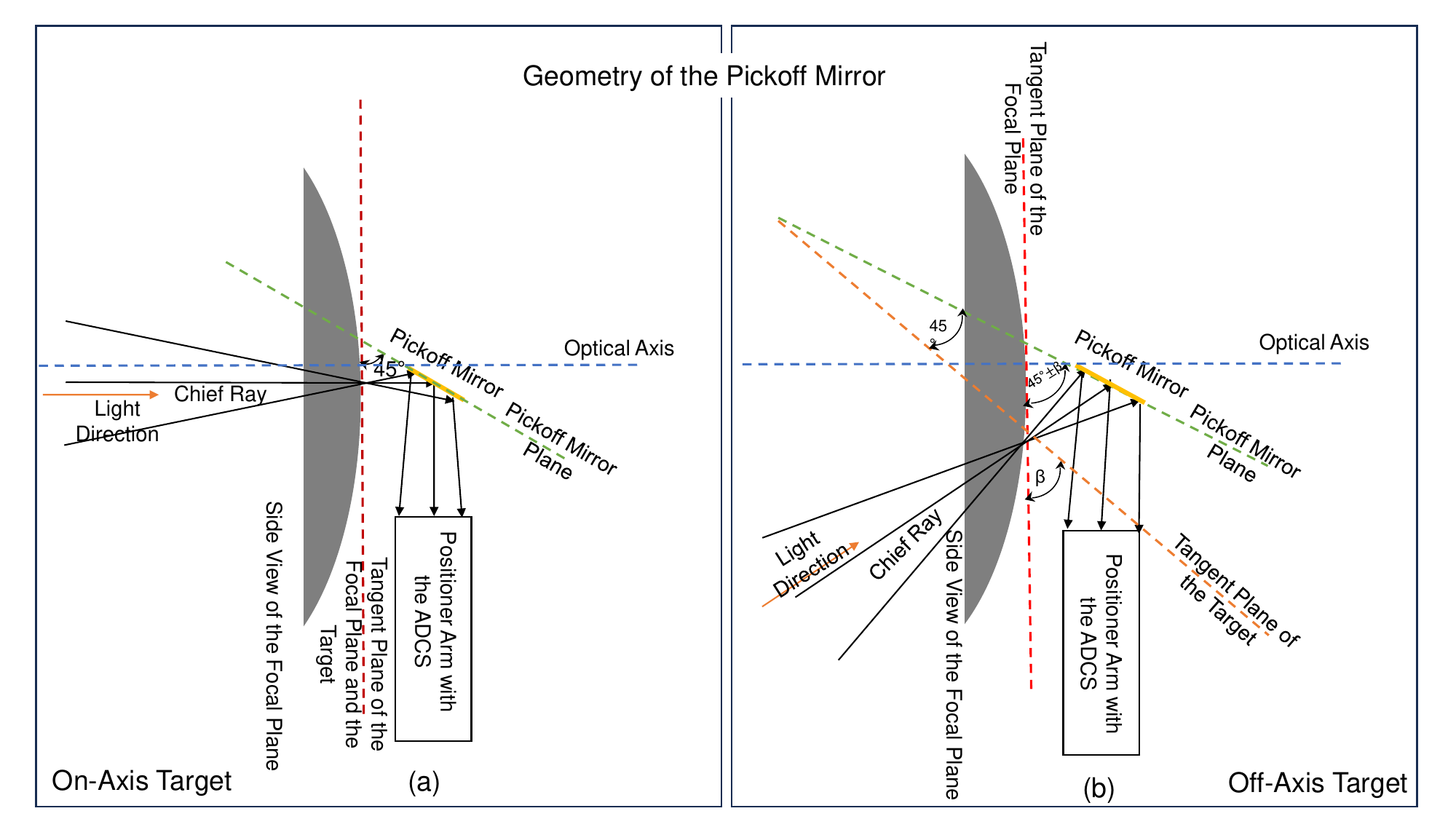}}
    \caption{(a) When the target aligns with the telescope's optical axis (on-axis), the pick-off mirrors (depicted by solid yellow lines) are oriented at an angle of 45\degree \ about the tangent of the target's chief ray, which redirects the light into the positioner arm which hosts the ADCS. (b) Conversely, when the target does not align with the telescope's optical axis (off-axis), the pick-off mirrors are adjusted to an angle of 45\degree \(\pm\) \(\beta\) to guide the light to the ADCS.}
    \label{gpm}
\end{figure}

The positioner's design encompasses an atmospheric dispersion corrector (ADC), fore optics such as a collimator, camera, dichroic elements, and a pick-off mirror. This collective assembly is termed the Atmospheric Dispersion Correction System (ADCS), and each positioner arm contains its own ADCS. The role of the pick-off mirror is to redirect non-telecentric light toward the collimator using tip and tilt motion. The mirror will maintain an angle of 45\degree\ with the chief ray of an on-axis target. For off-axis targets, the mirror will maintain an angle equal to 45\degree \(\pm\) the angle between the tangent plane of the on-axis object and the tangent plane of the off-axis object (\(\beta\)) \cite{Manju_2023}, as shown in Figures \ref{gpm}a and \ref{gpm}b. The collimated beam then passes through the ADC. The dispersion-corrected light is split into two wavelength bands, blue and red, using a dichroic and focused on optical fibers using a camera. These fibers send light to two channels of the spectrograph.

Here, the curved focal plane is defined in the spherical coordinate system. Consequently, the targets focused on this plane will also be in spherical coordinates, denoted as \( r_{t} \), \( \theta_{t} \), and \( \phi_{t} \). However, in this concept, the pick-off mirror diverts light into the positioner arm, which always remains perpendicular to the optical axis. Since the pick-off mirror ensures the light is perpendicular to the optical axis during the observation of the targets, the targets on the focal plane can be described using cylindrical coordinates denoted as \( r_{t} \), \( \theta_{t} \), and \( z_{t} \).

\begin{figure}[h!]
    \centering
    \subfigure{\includegraphics[width=1.01\textwidth,trim=0pt 20pt 0pt 20pt,clip]{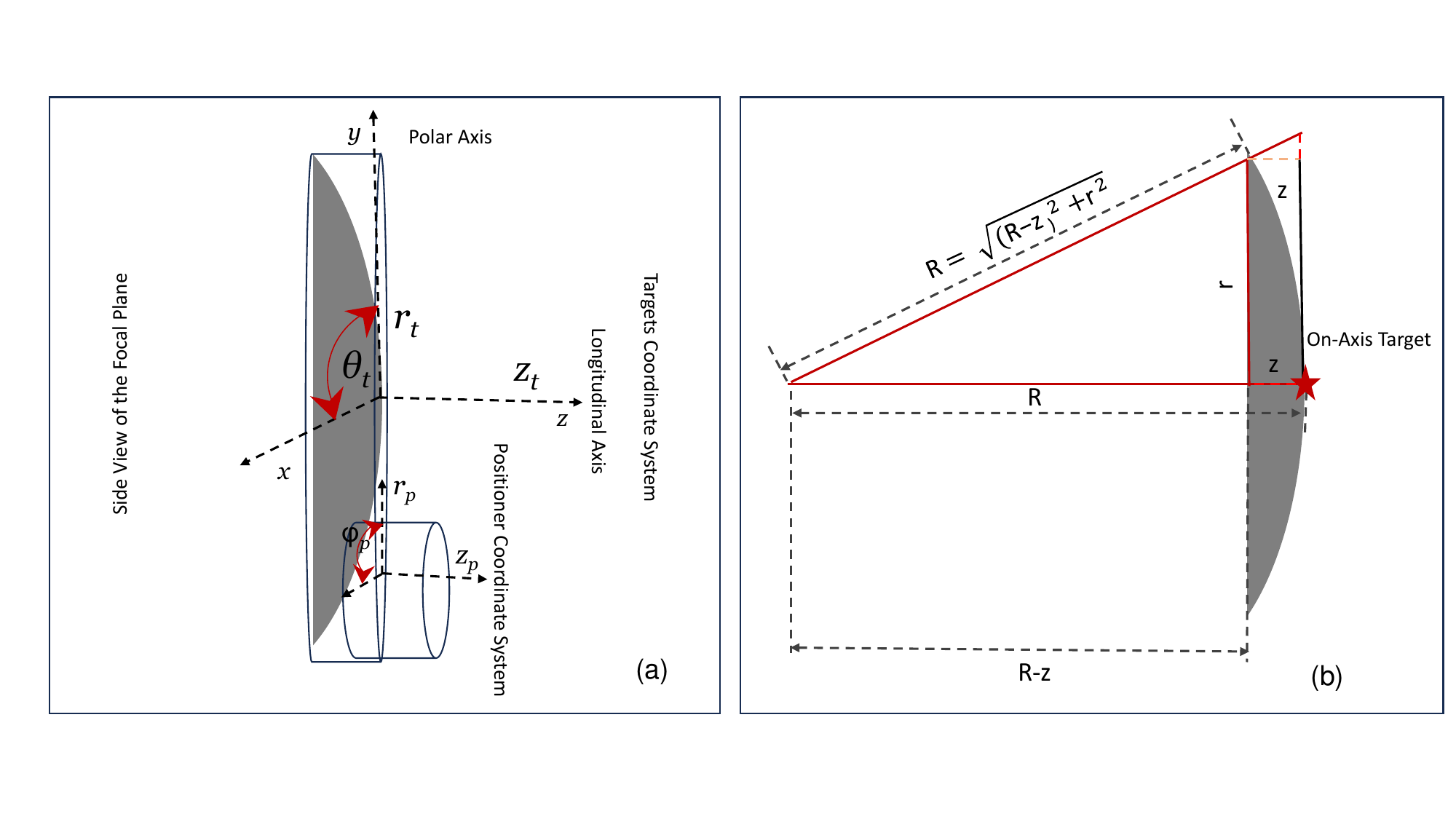}}
    \caption{(a) Target and positioner cylindrical coordinate system. (b) Relation between \(z\), \(r\), and \(R\).}
    \label{zrR}
\end{figure}

To place the positioner ferrule at any target angular position \( \theta_{t} \) on the focal plane of radius \( r_{f} \), the positioner base has to displace an angle \( \theta \) with respect to the x-axis about the optical axis or longitudinal axis of the focal plane. This angle varies from 0\degree\ to 360\degree\ on the periphery of the focal plane to reach the target's \( \theta_{t} \). Here, \( \theta \) is \( \theta_{p} - \theta_{t} \) (\( \theta_{p} \) is the positioner base angular location with respect to the x-axis of the target's coordinate system). Some targets might have almost the same or nearby \( \theta_{t} \) but different \( r_{t} \). To access such targets, which are close to or under the positioner arm, the bases of the positioners have to park at a certain angular distance apart from \( \theta_{t} \). Then, positioners ferrule can reach the targets by rotating the positioner arm about its longitudinal axis with an angle \( \phi_{p} \) to align with \( \phi_{t} \), which varies from 0\degree\ to 180\degree\ (where \( \phi_{t} \) represents the angle created by the positioner arm and the line extending from the base to the target), adjusting the ferrule \( r_{p} \) to \( r_{t} \) in the radial direction, and \( z_{p} \) to \( z_{t} \) longitudinally, as explained in Section \ref{section4.9}. Here, \( r_{p} \), \( \phi_{p} \), and \( z_{p} \) are the radial, azimuthal, and longitudinal coordinates of the positioner ferrule in cylindrical coordinates, as shown in Figure \ref{zrR}a.

Hence, the design of the positioner includes four degrees of freedom that help the positioner ferrule reach the target by additional linear and angular movements in addition to the movement of the pick-off mirror, as explained below.

\begin{enumerate}
    \item The positioner can revolve around the focal plane (denoted as \(\theta\), which is \( \theta_{p} - \theta_{t} \)), as shown in Figure \ref{Dof}a.
    \item The positioner arm can make an angular motion by rotating about its axis (denoted as \(\phi\), which is \(\phi_{t} \)), as depicted in Figure \ref{Dof}b.
    \item The positioner arm can move radially in and out (denoted as \(r\), which is \( r_{p} - r_{t} \)), as shown in Figure \ref{Dof}c.
    \item The positioner arm can move up and down in a longitudinal direction (denoted as \(z\), which is \( z_{p} - z_{t} \)), as shown in Figure \ref{Dof}d.
\end{enumerate}

Note that \(z\) and \(r\) depend on each other, as shown in Figure \ref{zrR}b.

\begin{equation*}
    R = \sqrt{(R - z)^2 + r^2}
\end{equation*}

\begin{equation*}
    (R - z)^2 = R^2 - r^2
\end{equation*}

\begin{equation}
    z = R - \sqrt{R^2 - r^2}
\end{equation}

Where \(R\) is the radius of curvature of the focal plane.

\newpage
\section{Conditions for Collision Avoidance Algorithm}
\label{section2}
\begin{figure}[h!]
    \centering
    \subfigure{\includegraphics[width=1\textwidth]{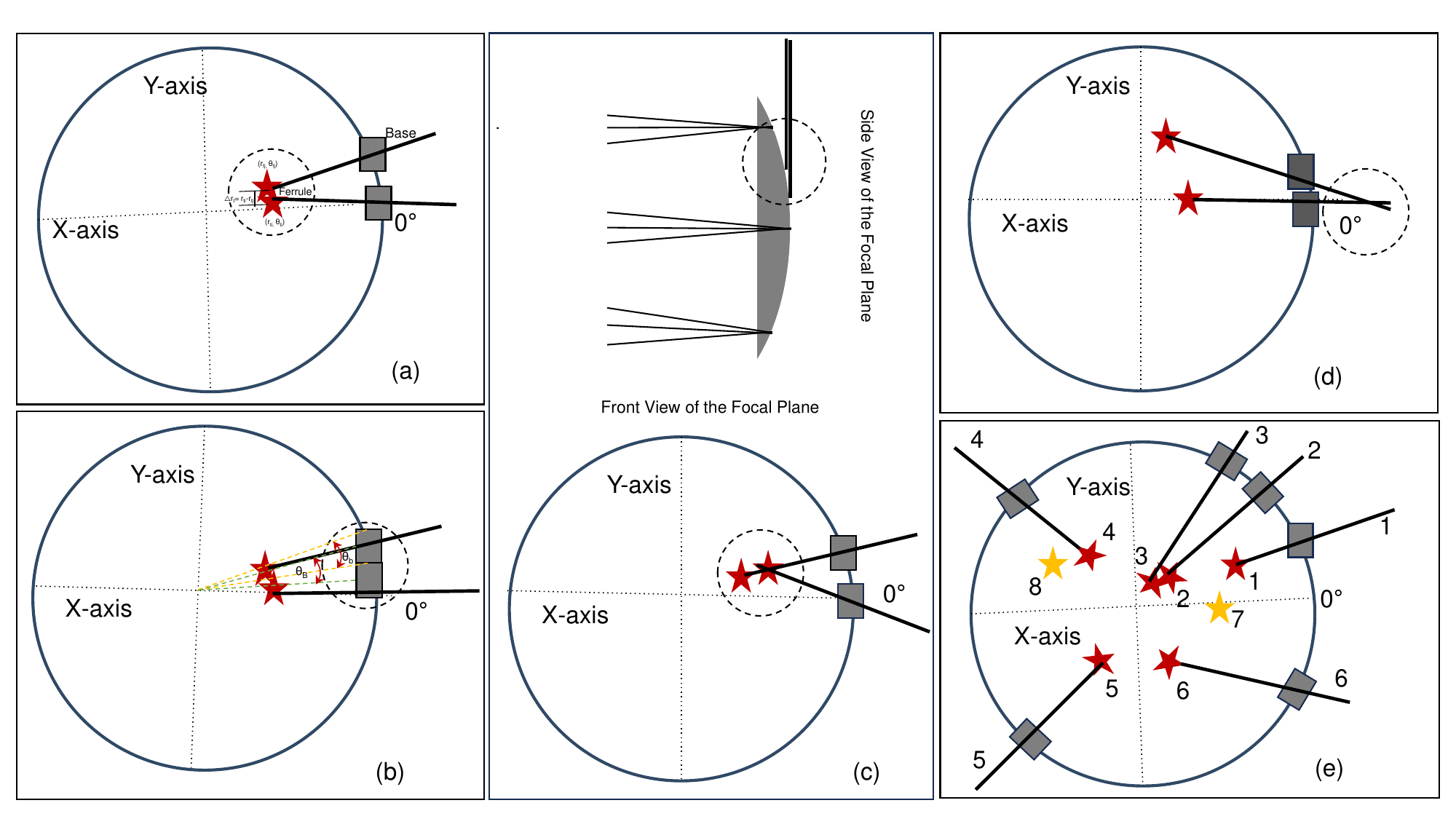}}
    \caption{The picture shows possible collision situations. Red stars represent the targets, black dotted circles mark potential collision zones, and gray boxes denote the bases. In image (e), yellow stars indicate targets not observed in the first session and will be observed in the next session.}
    \label{collisions}
\end{figure}

Having the appropriate assignment algorithm and motion planning of the positioner is crucial for better targeting efficiency and collision-free movement of positioners. Therefore, we assume that target assignment and motion planning can be effectively done under five main conditions:

\begin{enumerate}
    \item The distance between targets should be greater than the safe distance $d$ (see equation \ref{distance}), as shown in Figure \ref{collisions}a. This means the positioner's ferrule should not come too close to another ferrule.
    \item The angular separation between base positions concerning the center of the focal plane ($\theta_{B}=\theta_{pi}-\theta_{pj}$, where \( i \neq j \) and \( i, j = 1, 2, \ldots, n \)), should be greater than the angle covered by a base on the perimeter of the focal plane ($\theta_{b}$) (i.e., $\theta_{B} > \theta_{b}$), as shown in Figure \ref{collisions}b.
    \item Maintain the distance between two arms of the positioners greater than the safe distance $d$ when one arm passes beneath or over another to avoid a collision, as shown in Figure \ref{collisions}c.
    \item This assumption may not be required for the positioners arm operating with telescopic mechanisms. If not, then to prevent the arms from colliding outside the focal plane after a rotation by $\phi$, if targets are in the same plane (i.e., $\Delta r_{t} < d$, where $\Delta r_{t}$ is $r_{ti} - r_{tj}$ with $r_{ti}$ and $r_{tj}$ being the radial coordinates of the targets, and $i \neq j$, $i, j = 1, 2, \ldots, n$), as shown in Figure \ref{collisions}d.

    \item Avoid collisions during secondary or subsequent positioning sessions, i.e., during the movement of the positioner from one target to another (see section \ref{section4.9}).
\end{enumerate}

\newpage

\section{Methodology for Target Assignment and Collision Avoidance in Positioner Systems}

The procedure for efficiently assigning targets to positioners while minimizing collisions involves a detailed methodology that ensures accurate spatial representation of targets, optimal target assignment, and collision-free movement of positioners. Below is a step-by-step explanation of the process:

\subsection{Convert Equatorial Coordinates to Standard Coordinates}

The first step involves converting celestial coordinates, Right Ascension (RA) and Declination (Dec), to Cartesian coordinates using the following equations\cite{2005haip.book.....B}:

\begin{gather}
   X = \frac{\cos(\delta) \sin(\alpha - \alpha_{0})}{\cos(\delta_{0}) \cos(\delta) \cos(\alpha - \alpha_{0}) + \sin(\delta) \sin(\delta_{0})} \quad \text{radian}
\end{gather}

\begin{gather}
   Y = \frac{\sin(\delta_{0}) \cos(\delta) \cos(\alpha - \alpha_{0}) - \cos(\delta_{0}) \sin(\delta)}{\cos(\delta_{0}) \cos(\delta) \cos(\alpha - \alpha_{0}) + \sin(\delta) \sin(\delta_{0})} \quad \text{radian}
\end{gather}

Here, $\alpha$, $\delta$, $\alpha_{0}$, and $\delta_{0}$ represent the RA and Dec of an object and the center object in the field, respectively.

These Cartesian coordinates in radians are then converted to meters using the equations:

\begin{gather}
    x_{t} = \left( \frac{X \times S \times P}{1000} \right) \quad \text{meters}
\end{gather}

\begin{gather}
    y_{t} = \left( \frac{Y \times S \times P}{1000} \right) \quad \text{meters}
\end{gather}
% \begin{gather}
%      r_{t} = \sqrt{x_{t}^2 - y_{t}^2}
% \end{gather}
% \begin{gather}
%      z_{t} = R - \sqrt{R^2 - r_{t}^2}
% \end{gather}

Here, $P$ is the telescope plate scale in mm per arcsecond, and $S$ is the conversion factor from radians to arcseconds, defined as:

\begin{gather}
    S = \frac{3600 \times 180}{\pi}
\end{gather}

\textbf{Note:} Although in some sections like \ref{section1} and \ref{section4.9}, variables such as \(\theta\), (\( r_{t} \), \( \theta_{t} \), \( z_{t} \)), (\( r_{p} \), \( \phi_{p} \), \( z_{p} \)), \( \phi_{t} \), etc., are mentioned for individual targets and positioners, in subsequent sections, these variables are used collectively for group of targets and positioners, as follows:

\[
\theta = \{\theta_{p1}-\theta_{t1}, \theta_{p2}-\theta_{t2}, \ldots, \theta_{pn}-\theta_{tn}\} = \{\theta_{1}, \theta_{2}, \ldots, \theta_{n}\}
\]

\[
(r_{t}, \theta_{t} , z_{t}) = \{(r_{t1}, \theta_{t1}, z_{t1}), (r_{t2}, \theta_{t2}, z_{t2}), \ldots, (r_{tn}, \theta_{tn}, z_{tn})\}
\]

\[
(r_{p}, \phi_{p}, z_{p}) = \{(r_{p1}, \phi_{p1}, z_{p1}), (r_{p2}, \phi_{p2}, z_{p2}), \ldots, (r_{pn}, \phi_{pn}, z_{pn})\}
\]

\[
\phi_{t} = \{ \phi_{t1},\phi_{t2}, \ldots, \phi_{tn}\}
\]

\subsection{Conversion to Cylindrical Coordinates}

Next, each target's $x_{t}$, $y_{t}$ and $z_{t}$ coordinates are converted to cylindrical coordinates ($r_{t}$,$\theta_{t}$,$z_{t}$) using the following formulas: 

\begin{align}
    r_{t} &= \sqrt{x_{t}^2 + y_{t}^2}
\end{align}

\begin{align}
    \theta_{t} &= \arctan2(y_{t}, x_{t})
\end{align}

\begin{align}
    z_{t} &= R - \sqrt{R^2 - r_{t}^2}
\end{align}

\subsection{Sorting by Cylindrical Coordinates}

In a list, the targets are arranged in ascending order based on their angular coordinates $\theta_{t}$. If the difference between the angular coordinates of any targets, denoted as \( \Delta\theta_{t} = \theta_{ti} - \theta_{tj} \), where \( i \neq j \) and \( i, j = 1, 2, \ldots, n \), is less than \( \theta_{b} \), then consider them to have the same angular coordinate. Here, \( \theta_{ti} \) and \( \theta_{tj} \) represent the angular coordinates of the targets. These targets are sorted by their radial coordinates $r_{t}$. This method ensures the sequential assignment of targets to positioners and reduces the risk of base collisions.

\label{sec4.3}

\subsection{Distance Verification and Adjustment}

To avoid collisions between the ferrules and arms of positioners, it's important to maintain a minimum separation, denoted as d, between neighboring targets. Therefore, we calculate the Euclidean distance among all targets using their \( r_{t} \), \( \theta_{t} \), and \( z_{t} \) coordinates:

\begin{gather}
d_{ij} = \sqrt{(r_{ti} \cos \theta_{ti} - r_{tj} \cos \theta_{tj})^2 + (r_{ti} \sin \theta_{ti} - r_{tj} \sin \theta_{tj})^2 + (z_{ti} - z_{tj})^2}, \quad \text{for } i \neq j \text{ and } i, j = 1, 2, \ldots, n
\label{distance}
\end{gather}

If the distance $d_{ij}$ is less than a specified threshold dd, the target is moved to the end of the list.

\subsection{Calculation of Target-to-Positioner Ratio}

The value N = $\frac{T}{P}$ is calculated. If $N\neq 1$, then the value \(N = \left\lfloor \frac{T}{P} \right\rfloor + 1\) is determined, where \(T\) is the total number of targets and \(P\) is the number of available positioners. Sublists are created by splitting the primary list into N parts (where N is initially the number of observation sessions needed).

\subsection{Collision Avoidance within Sublists}

Within each sublist, the distances between targets are verified. If any two targets are closer than distance \(d\), the closer target is removed, and an attempt is made to place it in another sublist that meets the distance requirement and has fewer targets than the number of positioners. If no suitable sublist exists, a new one is created, and the target is added. This ensures that all sublist targets are adequately spaced to prevent collisions.

\subsection{Optimization of Observations}

Initially, sublists are sorted in order of increasing length (i.e., number of targets within a sublist). Suppose there are fewer targets than positioners in the early sublists. In that case, targets from neighboring shorter sublists are shifted to longer ones to fill gaps, as long as the required distance is maintained and the sublist doesn't exceed the positioner count. This approach ensures the maximum number of targets is observed at the start of the session. Sublists without any targets are then eliminated. The total count of remaining sublists represents the number of observation sessions needed to cover all targets ($N'$). This step optimizes observing more targets per session and reduces the overall number of sessions while maintaining safe distances to prevent collisions.

\subsection{Assign Targets to Positioners}

\begin{figure}[h!]
    \centering
     \subfigure{\includegraphics[width=1\textwidth, trim=0pt 150pt 0pt 120pt, clip]{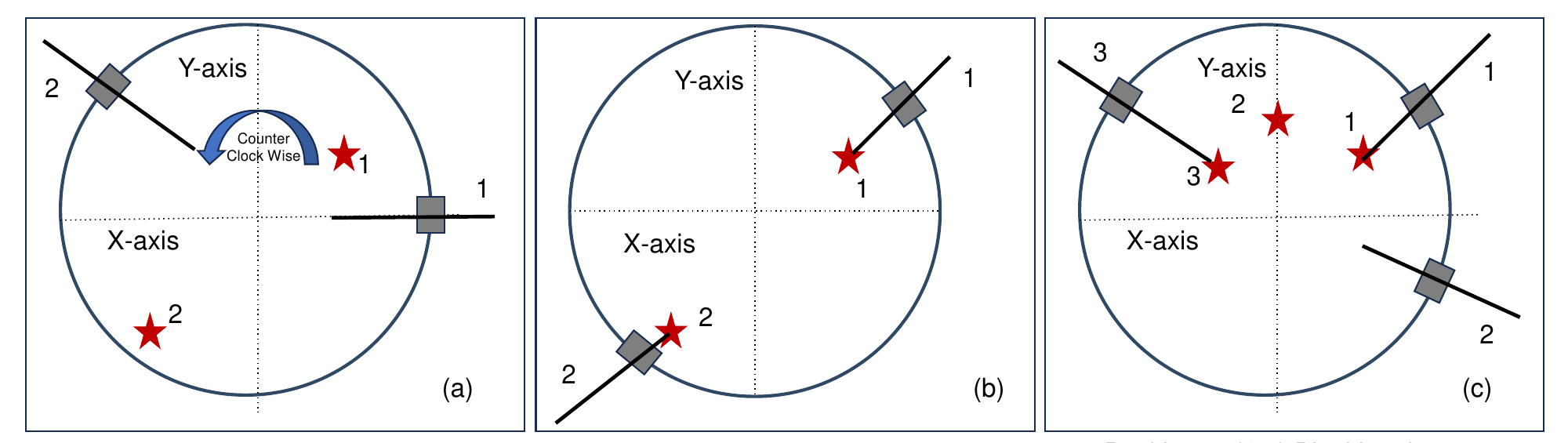}}
\caption{Target Assignment and Positioner Placement: (a) Positioners at their starting positions. (b) Positioners have moved in a counterclockwise direction to reach their respective targets. (c) Positioners 1 and 3 obstruct the path of positioner 2 to its target.}
    \label{final image}
\end{figure}

After the targets are segregated into sublists, each sublist is assigned to an observation session. The first target in the first sublist (i.e., the primary observation session) is selected and assigned to the nearest positioner. The remaining targets are then assigned to the remaining positioners in a clockwise or counterclockwise direction, based on the shortest direction the first positioner would take to move to its target, as shown in Figure \ref{final image}a \& \ref{final image}b. The choice of selecting the first target in the sublist can be random, but the assignment should be sequential, as the positioners must move around the focal plane. This sequential assignment prevents a scenario in which positioners block each other. For instance, consider a scenario with three targets and three corresponding positioners labeled 1, 2, and 3. The first and third targets are assigned to positioners 1 and 3. However, the second target is located between the first and third positioners. If not assigned sequentially, this could obstruct the second positioner from accessing its intended target, as shown in Figure \ref{final image}c.

Since positioners are assigned sequentially, they move around the focal plane one by one. However, secondary or subsequent observational sessions are required if $N' > 1$. In these sessions, the target assignment has to take place with respect to the new positions of the positioners because their positions change at the end of the previous observation session.

\label{sec4.8}

\subsection{Place the Positioners ferrule on the Targets}

In any session, if the number of targets in the sublist, denoted as \(s\), is less than the number of positioners, \(p\), there is a possibility that unassigned positioners in the ongoing observation session could collide with those that are assigned. For example, consider a scenario with eight targets (numbered 1 to 8) and six positioners (numbered 1 to 6). The value of \(N'\) is 2, which satisfies the condition \(N' > 1\). In this case, the first sublist will contain six targets, and the second sublist will contain two targets, thus satisfying the condition \(s < p\). In the initial observation session, six positioners observe six targets. As per the methodology, the 7th target is assigned to the 1st positioner in the subsequent observation session. The 8th target is assigned to the positioner that comes after the first one in a clockwise direction. So, the 8th target goes to the 6th positioner. However, the 5th positioner is already at the 5th target, so for the 6th positioner to reach the 8th target, it has to cross the path of the 5th positioner, which is not assigned in the second observation session. This could lead to a collision between the 5th and 6th positioners if the 5th positioner is not moved to a safe position, as illustrated in Figure \ref{collisions}e.

To avoid this, dummy targets are generated with a fixed radial coordinate (\(r_{f}\)) and varying angular coordinates (\(\theta_{r}\)) using the Monte Carlo method. These dummy targets are inserted into the sublist, provided the distance condition is met, and the sublist does not contain more targets than the number of positioners. Subsequently, the targets in the sublist are sorted and assigned to positioners, as detailed in Sections \ref{sec4.3} and \ref{sec4.8}.

After assigning the targets, the positioners move from their initial \(\theta_{p}\) to the targets \(\theta_{t}\) by adding \(\theta\) to their positions. All other positioners follow the direction of the first positioner, clockwise or counterclockwise, to park their bases at the final positions \(\theta_{pf}\), which should be \(\theta_{pf} = \theta_{t}\) if the condition \(\theta_{B} > \theta_{b}\) is satisfied among all the positioner bases. If this condition is not met, the \(\theta_{pf}\) of the targets is adjusted by \(\Delta \theta_{s}\) until the condition is satisfied. The updated \(\theta_{pf}\) is denoted \(\theta_{pf'}\), which changes \(\theta_{B}\). Then, \(r\), \(\phi_{t}\), and \(z\) are recalculated for the new \(\theta_{pf} \pm \Delta \theta_{s}\) to position the ferrule on the target. The updated \(r\), \(\phi_{t}\), and \(z\) are indicated as \(r_{rec}\), \(\phi_{rec}\), and \(z_{rec}\), respectively. Finally, the ferrules are positioned on the target by adding \(r_{rec}\), \(\phi_{rec}\), and \(z_{rec}\) to the updated ferrule coordinates \(r_{p}\), \(\phi_{p}\), and \(z_{p}\).

However, rotation of \(\phi_{rec}\) could lead to a collision, as mentioned in point 3 in Section \ref{section2}. To avoid this, the positioner arms are modeled as three-point line segments defined by points at the ferrule, midpoint, and endpoint, corresponding to the arm's length. Any points of contact between adjacent line segments during each \(\phi_{rec}\) rotation are checked. If a point of contact is detected, the positioner angular position on the focal plane is adjusted as mentioned above, changing \(\theta_{pf'}\) by \(\Delta \theta_{s}\), resulting in new coordinates \(r_{rec'}\), \(\phi_{rec'}\), and \(z_{rec'}\) that define a new line segment. This adjustment is repeated until there are no points of contact between line segments, ensuring that the arms do not collide during movements. Finally, the ferrules are positioned on the target by adding \(r_{rec'}\), \(\phi_{rec'}\), and \(z_{rec'}\) to the updated ferrule coordinates \(r_{p}\), \(\phi_{p}\), and \(z_{p}\).

% \section{Quantifying the Algorithm}

% To evaluate the algorithm, the Monte Carlo technique is used. Fifty targets are randomly picked from a pool of 5000 without any restrictions on magnitude, and 50 positioners are selected. The algorithm is run 10000 times, and the proportion of targets that could be observed in each run during the initial observation session is calculated; see Figure \ref{Results} a.

% To determine the number of observation sessions required to observe all targets, the sessions needed to observe 50, 60, 70, 80, 90, and 100 percent of the targets in each run are tracked. The average results for each percentage over the 10000 runs are then taken; refer to Figure \ref{Results}b.

\label{section4.9}

\section{Quantifying the Methodology}

\begin{figure}[h!]
    \centering
    \subfigure{\includegraphics[width=1\textwidth, trim=0pt 80pt 0pt 60pt, clip]{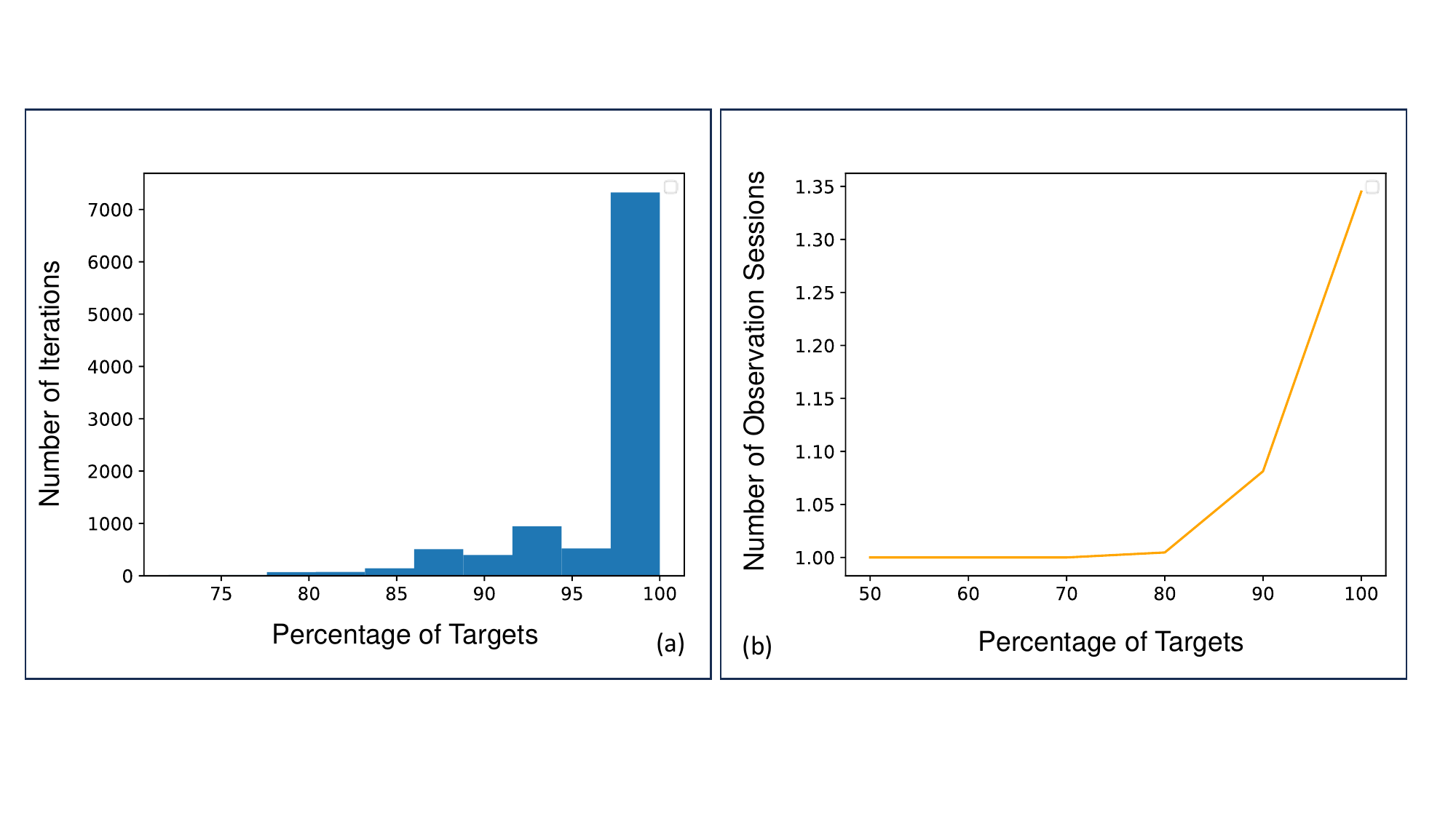}}
    \caption{(a) a histogram where the x-axis represents the number of iterations and the y-axis represents the percentage of targets observed in the first observation session. (b) the average number of observation sessions needed to observe 50\%, 60\%, 70\%, 80\%, 90\%, and 100\% of the targets over 10,000 iterations.}
    \label{Results}

\end{figure}

We used the Monte Carlo method to assess the algorithm. From the Global Astrometric Interferometer for Astrophysics (GAIA) catalog, we randomly selected 50 from around 5,000 targets for a 20' field centered at RA, Dec is 280\degree,-60\degree, with no magnitude limitations, and chose 50 positioners (N = 1). The algorithm was run 10,000 times, considering the parameter values given in Table\ref{table1}, and we calculated the proportion of targets that could be observed in each run during the primary observation session. It was found that in 8,000 iterations, 100\% of the given targets could be observed in the first session, as shown in Figure \ref{Results}a

To determine the number of observation sessions required to observe all targets, we tracked the sessions needed to observe 50\%, 60\%, 70\%, 80\%, 90\%, and 100\% of the targets in each run. We then averaged the results for each percentage over the 10,000 runs. The result was that 80 percent of the targets could be observed in the primary observation session and the remaining 20 percent in the secondary observation session. In total, two observation sessions were sufficient to observe all targets (N' = 2), as depicted in Figure \ref{Results}b.

% \section{Quantifying the Algorithm}

% To evaluate the algorithm, we used the Monte Carlo technique. We randomly picked 50 targets from a pool of 5000, without any restrictions on magnitude, and selected 50 positioners. We ran the algorithm 10000 times and calculated the proportion of targets that could be observed in each run during the initial observation session; see Figure \ref{Results} a

% To determine the number of observation sessions required to observe all targets, we tracked the sessions needed to observe 50, 60, 70, 80, 90, and 100 percent of the targets in each run. We then took the average of the results for each percentage over the 10000 runs refer Figure \ref{Results}b 
.

\begin{table}[h!]  
    \caption{Considered specifications for quantifying the methodology} % title name of the table  
    \centering % centering table  
    \begin{tabular}{p{7cm}p{2cm}} % creating 2 columns 
    \hline\
    \centerline {Parameters} & Values \\ [0.5ex]   
    \hline\
     R & $\approx$ 3m\\
     Field of view & 2.56m (20') \\
     T$\&$P & 50 \\
     $\alpha_{0}$, $\delta_{0}$& 280\degree,-60\degree \\
     length of arm & 1.2m \\
     $\theta_{b}$ & 5\degree \\
     $\triangle \theta_{s}$ & 0.01\degree \\
     d & 0.05m \\
     Number of Monto Carlo iterations & 10,000 \\

    % [1ex] adds vertical space  
    \hline % inserts single-line  
    \end{tabular}
    \label{table1}
\end{table}

% Figure \ref{Results}

% Figure \ref{Results}

\section{Conclusion and Future Work}

The proposed fiber positioning system, designed for multiple purposes with four degrees of freedom, faces challenges due to its long arm length of approximately 1.2 meters, which is necessary to reach the on-axis target from the focal plane's edge. The need for four motors to facilitate the four degrees of freedom results in a larger base size, making the system only suitable for medium-scale multiplexing. Although systems with fewer degrees of freedom are less prone to failure, this design has potential if adequately balanced. Our Python algorithm, designed for T$\times$P targets and positioners for target assignment and positioning, demonstrates that all targets can be covered in 80$\%$ iteration during the first session. On average, all targets are observed within two sessions.
Moreover, de-rotating the field rotator, where the entire FPS is located, can alleviate stress caused by field rotation on fibers. This concept also eliminates the need for large atmospheric dispersion for a wider field of view and compensates for non-telecentricity. Therefore, we believe this concept is a promising choice for specific targeting and surveying using seeing-limited multi-object high-resolution spectrographs on large telescopes where precision is not a top priority.

In the future, we will develop the algorithm for actual mechanical design parameters and plan to fine-tune it to lessen the steps needed. In addition, we will run simulations to understand the trade-off between precision in positioning and throughput loss. And how changes in the stability of positioners during the observation session affect radial velocity. We will also work on selecting the best mechanical parts to achieve our desired precision and stability.

 \begin{figure}[h!]
    \centering
    \subfigure{\includegraphics[width=1\textwidth]{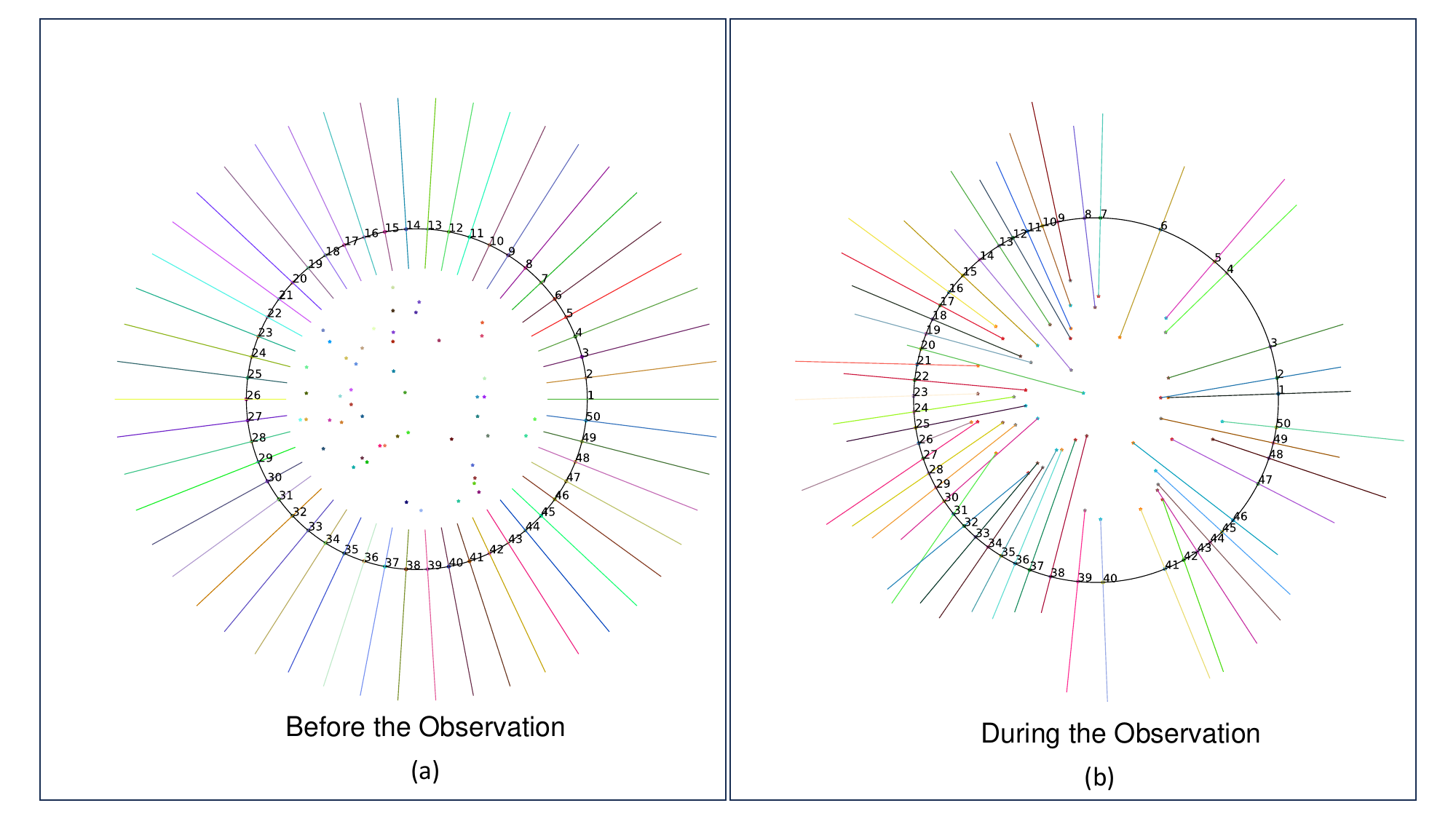}}
    \caption{In these 2D images, positioners and targets are depicted on the focus plane (a) Positioners parked at their designated locations before observation, (b) In this image, the positioners seem to overlap because they are depicted on a flat plane. However, in three-dimensional space, the positioners are spatially separated and accurately positioned on their targets during observation.}
    \label{rpt}

\end{figure}

\newpage
%\conclusion
\acknowledgments % equivalent to \section*{ACKNOWLEDGMENTS}  

Manjunath would like to thank S. R. Dhanush for his thoughtful conversations and recommendations. In addition, he expresses gratitude to the Wide-Field Optical Spectrometer (WFOS-TMT) team for providing the conceptual structure design, which was crucial in integrating the FPS concept. We sincerely thank the Indian Institute of Astrophysics and the Department of Science and Technology for their constant support throughout this project.
 
% References

% References
\bibliography{report} % bibliography data in report.bib

\bibliographystyle{spiebib} % makes bibtex use spiebib.bst

% \appendix    %>>>> this command starts appendixes

% \section{Pick off mirror}

% % \section{z}

%  \begin{figure}[h!]
%     \centering
%     \subfigure{\includegraphics[width=0.85\textwidth]{Imgaes/outputfile_page_7.pdf}}
%     \caption{Relation between z,r and R}
%     \label{zrR}

% \end{figure}

%  \begin{figure}[h!]
%     \centering
%     \subfigure{\includegraphics[width=0.85\textwidth]{Imgaes/outputfile_page_8.pdf}}
%     \caption{Geometry of the pick of mirror}
%     \label{gpm}

% \end{figure}

\end{document}